\magnification=\magstep1
\def\para{\par\noindent}
\def\sqr#1#2{{\vcenter{\vbox{\hrule height.#2pt
        \hbox{\vrule width.#2pt height#1pt \kern#1pt
          \vrule width.#2pt}
        \hrule height.#2pt}}}}
\def\square{\mathchoice\sqr34\sqr34\sqr{2.1}3\sqr{1.5}3}
\newcount\notenumber

\def\note{\advance\notenumber by 1
\footnote{$^{\the\notenumber}$}}
\baselineskip 20pt
\centerline{{\bf Scaling and Persistence in the}}
\centerline{{\bf Two-Dimensional Ising Model}}\para
\vskip 0.25cm
 \para S.~Jain$^{1}$ and H.~Flynn,
 \para School of Mathematics and Computing,\para
 University of Derby,\para
 Kedleston Road,\para
 Derby DE22 1GB,\para
 U.K.\para
\vskip 2.25cm
\para $^{1}$ Address after 17$^{th}$ April 2000: 
\para School of Engineering and Applied Sciences,\para
                    Aston University,\para
Aston Triangle,\para
Birmingham B4 7ET,\para
U.K.\para
\vskip 3.0cm
\para Classification Numbers: 
\para 05.20-y, 05.50+q, 05.70.Ln, 64.60.Cn, 75.10.Hk, 75.40.Mg 
\para
\vfill\eject
\para {\bf ABSTRACT}
\para The spatial distribution of persistent spins at zero-temperature 
in the pure two-dimensional
Ising model is investigated numerically. A persistence correlation length,
$\xi (t)\sim t^{Z}$ is identified such that for length scales
 $r <<\xi (t)$ the persistent
spins form a fractal with dimension $d_f$; for length scales $r >>\xi (t)$ the
distribution of persistent spins is homogeneous. The zero-temperature 
persistence exponent, $\theta$, is found to satisfy the 
scaling relation $\theta =
Z(2-d_f)$ with $\theta = 0.209\pm 0.002, Z=1/2$ and $d_f\sim 1.58$. 
\para 
\vfill\eject
\para The \lq persistence\rq\ problem has attracted considerable interest in
recent years [1-9]. In its most general form, it is concerned
 with the fraction of space
which persists in its initial state up to some later time.
\para Hence, in the non-equilibrium dynamics of spin systems
 at zero-temperature we
are interested in the fraction of spins, $P(t)$, that persist in the same state
as at $t=0$
up to some later time $t$.
 For the pure ferromagnetic two-dimensional Ising model, $P(t)$ has
been found to decay algebraically [1-4] 
$$P(t)\sim t^{-\theta } \eqno(1)$$
 where $\theta = 0.209\pm 0.002$ [5]. Similar algebraic decay has been found in
numerous other systems displaying persistence [9]. Most of the recent
theoretical effort has gone into obtaining the numerical value of $\theta$ for
different models.
\para Very recently, Manoj and Ray [10] have studied the spatial correlation
of persistent sites in the $1d\ A+A\rightarrow 0$ model. They found that
the set of
persistent sites in their $1d$ model forms a fractal over sufficiently small
length scales.
 
\para In this letter we present the results of an extensive numerical
 study of the spatial distribution of persistent spins in
 the
pure $2d$ Ising model at zero-temperature. As we will
 see, the $2d$ Ising model exhibits behaviour very similar to that 
found by Manoj and Ray [10]
  in their simple $1d$ model.
 
\para The Hamiltonian for our model is given by
$${\it H} = -\sum_{<ij>}  S_i S_j\eqno(2)$$
where $S_i=\pm 1$ are Ising spins situated on every site of a
 square lattice with periodic boundary conditions; the summation in Eqn. (2) 
runs
 over all nearest-neighbour pairs only.

\para The data presented in this work were obtained for a lattice with dimensions
 $1000\times 1000\ (=N)$.
\para Each simulation run begins at $t=0$ with a random
 ($\pm 1$) starting configuration 
of the spins and
then we update the lattice via single spin flip zero-temperature
 Glauber dynamics [5].
 The rule we use is: always flip if the energy
 change is negative, never flip if the energy change is positive and flip at
random if the energy change is zero.  
\para For each spin $S_i$ we define
$$n_i(t)=(S_i(t)S_i(0)+1)/2.\eqno(3)$$
Hence, if $n_i(t)=1$ for all $t\ge 0$ spin $S_i$ 
is persistent at time $t$; $n_i(t)=0$ otherwise.
\para The total number, $n(t)$,
of spins which have never flipped until time $t$ is then given by
$ n(t)=\sum_{i}n_i(t),$
and the persistence 
probability by [1] 
$$P(t)=\sum_{i}<n_i(t)>/N\eqno(4)$$
where $<\dots>$ indicates averages over different 
initial conditions and histories.
We averaged
over at least 100 different initial conditions and histories for each run.
\para To investigate the spatial correlations in this model, we follow Manoj and
Ray [10] and study the 2-point correlator defined by
$$ C(r,t) = <n_i(t)n_{i+r}(t)>/<n_i(t)>, \eqno(5)$$
where $<\dots>$ now also includes the average over the lattice shown explicitly
in Eqn (4).
$C(r,t)$ is simply the probability that spin $n_{i+r}(t)$ is persistent given
that $n_i(t)$ is persistent, averaged over the entire lattice. According to
 [10],
the 2-point correlator satisfies the following dynamic scaling relation
$$C(r,t) = P(t)f(r/\xi (t))\eqno(6)$$
where $\xi (t)$ is the persistence correlation length and $f(x)$ is a scaling
function such that 
$$f(x)\sim \cases{x^{-\alpha}, &for $x<<1$;\cr
                  1, &for $x>>1$.\cr}\eqno(7)$$  
As a consequence, the expected behaviour of $C(r,t)$ in the two limits is given
by
$$C(r,t)\sim \cases{r^{-\alpha} &for $r<<\xi (t)$;\cr
                    t^{-\theta} &for $r>>\xi (t)$.\cr}\eqno(8)$$
Clearly, as $P(t)\sim t^{-\theta}$, we must also have $\xi^{-\alpha}\sim t^{-\theta}$
to satisfy Eqn (8) in the limit $r<<\xi (t)$. Assuming a power-law
divergence for the persistence correlation length
 with $t$ i.e. $\xi (t)\sim t^Z$
then leads to the scaling relation $Z\alpha = \theta$. As we are working with
the pure $2d$ Ising model at zero-temperature, we expect [11] $Z=1/2$; our 
results are completely consistent with this assumption. 
\para To examine the correlated region ($r<<\xi (t)$) we study the average number
of persistent spins, $n(l,t)$, in a square grid with dimensions $l\times l$. As
$$n(l,t)=\int_{0}^{l}C(r,t)rdr\eqno(9)$$
we have that
$$ n(l,t)\sim \cases{l^{2-\alpha} &for $l<<\xi (t)$;\cr
                     l^2P(t) &for $l>>\xi (t)$.\cr}\eqno(10)$$  
Hence, we expect the persistent spins to form
 a fractal with dimension $d_f=2-\alpha$ for
length scales $l<<\xi (t)$; the distribution is homogeneous on longer length
scales, namely for $l>>\xi (t)$. We expect the crossover to occur
 at $l\approx\xi (t)\sim t^{1/2}$.
The scaling form for $n(l,t)$ is given by
$$ n(l,t) = l^2P(t)g(l/\xi(t)), \eqno(11)$$
where $g(x)$ is a scaling function satisfying
$$g(x)\sim \cases{x^{-\alpha} &for $x<<1$;\cr
                  1           &for $x>>1$.\cr}\eqno(12)$$
 We now discuss our results. 
\para Figure 1 shows a plot of the scaling function
$f(x)(=C(r,t)/P(t))$ against $x=r/\xi (t)$ for various different values of $t$.
 We have assumed that $\xi (t)\sim 
t^{1/2}$. The data in Fig 1 ranges over almost three orders of magnitude and
 is clearly consistent with this assumption. The large $x$ behaviour of $f(x)$
clearly follows the expected behaviour given in Eqn (7).
\para To extract a value for $\alpha$ we
re-plot the data shown in Fig 1 on a log-log scale in Fig 2. The algebraic 
behaviour for $x<<1$ of the scaling function is confirmed by the linear fit. The
slope of the straight line implies a value of $\alpha = 0.428\pm 0.007$. Hence,
the scaling relation would suggest that $\theta = Z\alpha = 0.214\pm 0.004$. 
This is, of course, consistent with value $(0.209\pm 0.002)$ quoted above
 for $\theta$ [5]. 
\para We investigate the correlated regions by obtaining a direct estimate
 of the fractal dimension $d_f$. This is undertaken by first partitioning the
lattice into square grids of size $l\times l$ with $l$ ranging from 4 to 250.
The average number of persistent spins in each $l\times l$ square is then
 obtained.
\para In Figure 3 we plot $\ln n(l,t)$ versus $\ln l$ for $t=10^2, 10^3,
5\times 10^3$ and $10^4$. We notice that for each of the values of $t$, the
behaviour over sufficiently small (typically, $l<<\sqrt t$) length scales is
consistent with a fractal dimension $d_f=2-\alpha\sim 1.58$; over longer
length scales (typically, $l>>\sqrt t$) we retrieve homogeneous behaviour
($d_f=d=2$). Actual values of
$d_f$ range from $d_f(t=10^2)\sim 1.62$ to $d_f(t=10^4)\sim 1.58$. The straight
lines, with slopes $1.58$ and $2.00$, shown in Fig 3 are linear fits to the
 behaviour in the two respective regimes for $t=10^4$.
\para We obtain an independent
estimate for the exponent $\alpha$ by re-plotting the data for $t=5\times 10^3$
and $10^4$ in scaling form. Figure 4 shows a log-log plot of the scaling function
$g(x)=n(l,t)/l^2P(t)$ against $x$ where $x=l/\sqrt t$. We see that the data
 clearly
fall onto a single scaling curve consistent with the expected behaviour given
in Eqn (12). On fitting all of the data for $\ln x < -0.5$ we get a value
of $\alpha\sim 0.438$. However, restricting the linear fit to $\ln x < -1$,
as indicated by the straight line in Fig 4, would imply a value 
of $\alpha\sim 0.50$. Although this is slightly higher than the value we obtained
from the analysis of the scaling behaviour of the 2-point correlator (see 
Eqn (8)), it is, nevertheless, consistent with our value of the fractal dimension
in the correlated regime.  
\para To conclude, we have investigated the spatial distribution of persistent
spins at zero-temperature in the pure two-dimensional Ising model. We find that
the persistent spins form a fractal with dimension $d_f\sim 1.58$ for length
scales $r<<\xi (t)$, where $\xi (t)\sim t^Z$ is the persistence correlation
length. Furthermore, the persistence exponent satisfies 
the scaling relation $\theta=Z(2-d_f)$ with $Z=1/2$. 
\para {\bf Acknowledgement}
\para The simulations were performed partly on the SGI Origin 2000 at the University
of Manchester made available by the Engineering and Physical Sciences Research
Council (EPSRC), Great Britain, and also on in-house workstations and a PC.
 HF would
like to thank the University of Derby for a Research Studentship
\vfill\eject 
\para FIGURE CAPTIONS
\para 
\vskip 1cm
\para Fig. 1
\para A plot of the scaling function $f(x) (=C(r,t)/P(t))$
 against $x$ where $x=r/\sqrt t$ for $t$ ranging over approximately three
orders of magnitude.
\vskip 1cm
\para Fig. 2
\para A re-plot of the data shown in Figure 1 on a log-log scale. The straight
line implies a value of $\alpha = 0.428\pm 0.007$.
\vskip 1cm
\para Fig. 3 
\para A log-log plot of $n(l,t)$ against $l$. Here, $n(l,t)$ is the
 average number of persistent spins in
a square ($l\times l$) grid at time $t$. The data is shown for (top) $t= 10^2\
 \diamondsuit, 
10^3\ +, 5\times 10^3\ \square$\ and $10^4\ \times$ (bottom). There is a
 clear crossover at $l\approx\sqrt t$ from a
 fractal distribution with dimension $d_f\sim 1.58$ to a homogeneous
one with $d_f=d=2$. The two straight lines (with slopes 1.58 and 2.00) 
are fits of the data in the two extreme cases for $t=10^4$. 
\vskip 1cm
\para Fig. 4
\para A plot of $\ln g(x)$ against $\ln x=\ln l/\xi (t)$. Here the scaling function
$g(x)=n(l,t)/l^2P(t)$. The straight line has slope $= -0.50$ and implies
a value of $\alpha\sim 0.50$.  
\vfill\eject
\para REFERENCES
\item {[1]} B. Derrida, A. J. Bray and C. Godreche, J.Phys. A {\bf 27},
 L357 (1994).
\item {[2]} A.J. Bray, B. Derrida and C. Godreche, Europhys. Lett. {\bf 27},
 177 (1994).
\item {[3]} D. Stauffer J.Phys.A {\bf 27}, 5029 (1994).
\item {[4]} B. Derrida, V. Hakim and V. Pasquier, Phys. Rev. Lett. {\bf 75},
 751 (1995); J. Stat. Phys. {\bf 85}, 763 (1996).
\item {[5]} S. Jain, Phys. Rev. E{\bf 59}, R2493 (1999).
\item {[6]} S.N. Majumdar, C. Sire, A.J. Bray and S.J. Cornell, Phys. Rev. Lett.
 {\bf 77}, 2867 (1996). 
\item {[7]} B. Derrida, V. Hakim and R. Zeitak, Phys. Rev. Lett. {\bf 77}
 2971 (1996).
\item {[8]} S.N. Majumdar and A.J. Bray, Phys. Rev. Lett. {\bf 81} 2626 (1998).
\item {[9]} S.N. Majumdar, {\it Curr. Sci.} {\bf 77} 370 (1999)
\item {[10]} G. Manoj and P. Ray, J.Phys. A{\bf 33}, L109 (2000)
\item {[11]} J.D. Gunton, M. San Miguel, and P.S. Sahni, {\it Phase Transitions
and Critical Phenomena}, edited by C. Domb and J.L. Lebowitz (Academic Press,
New York, 1983), vol 8
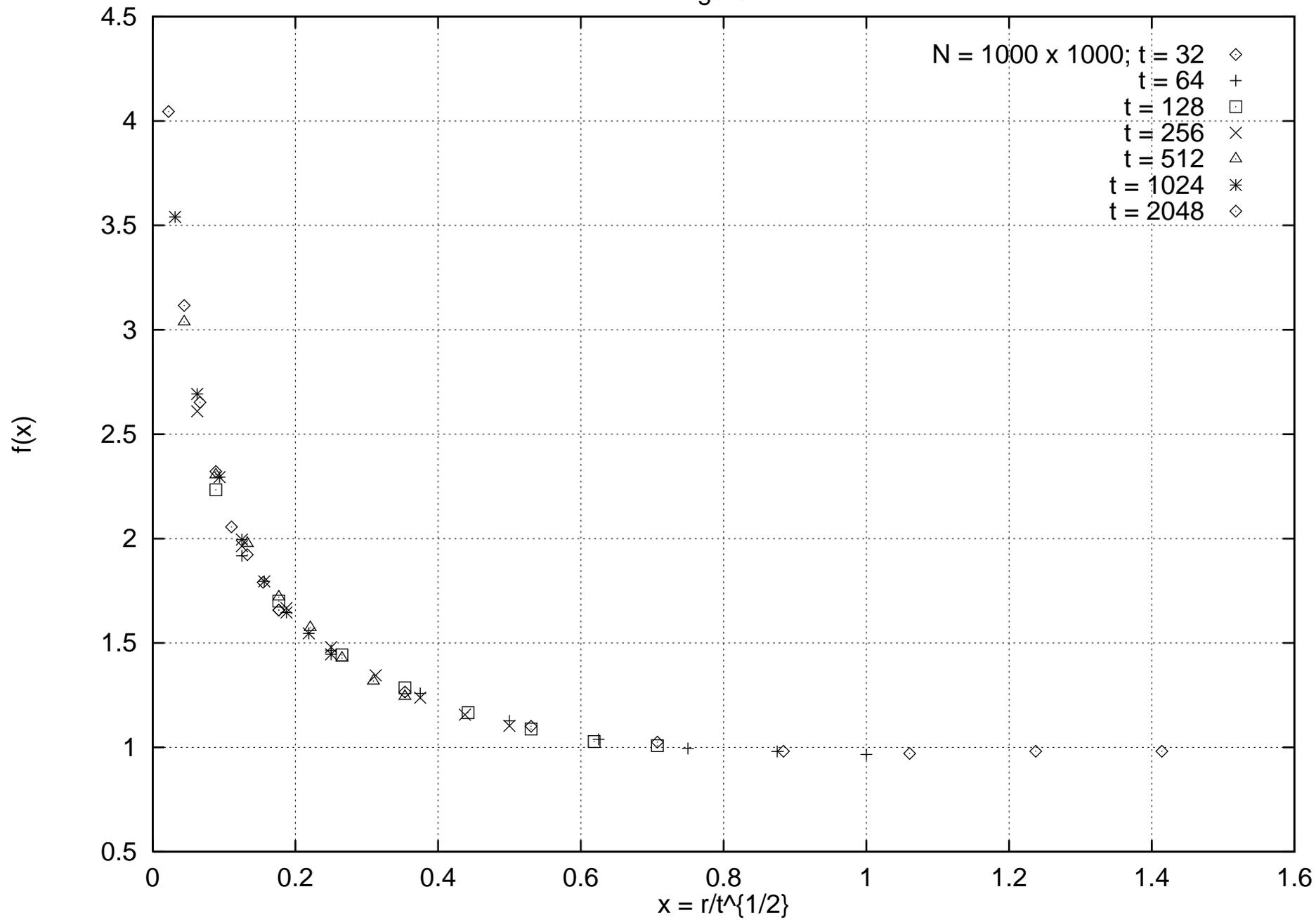
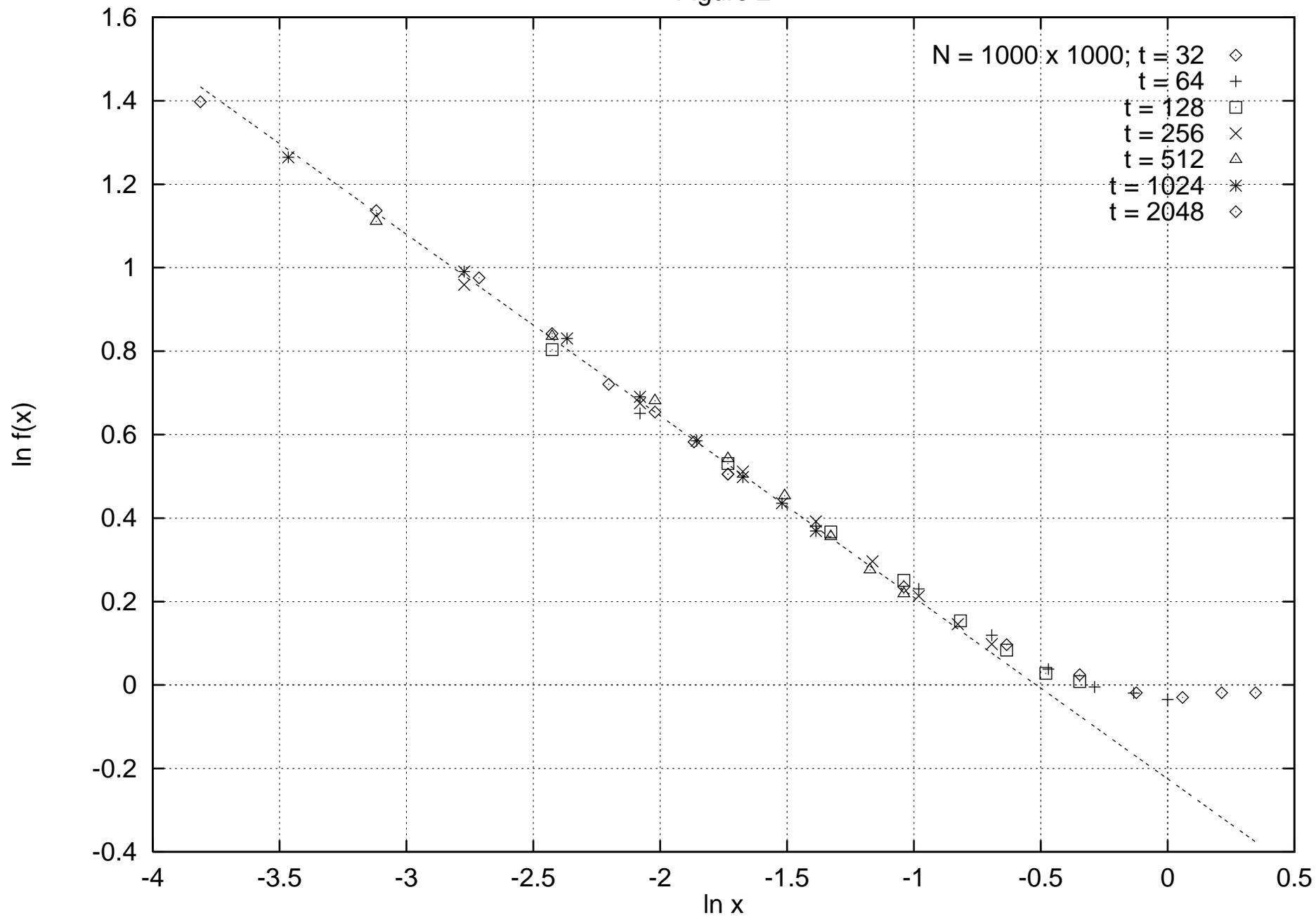
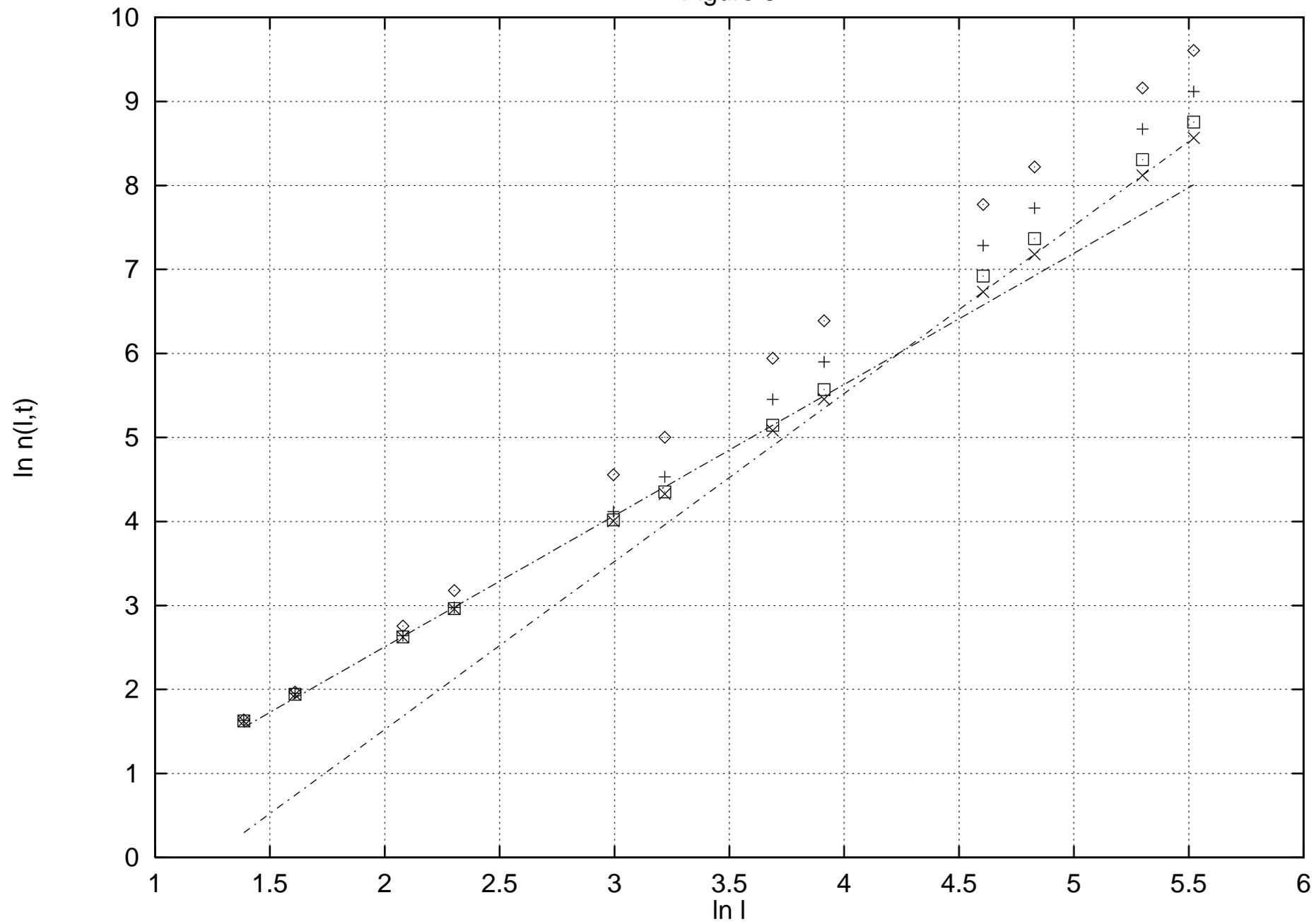
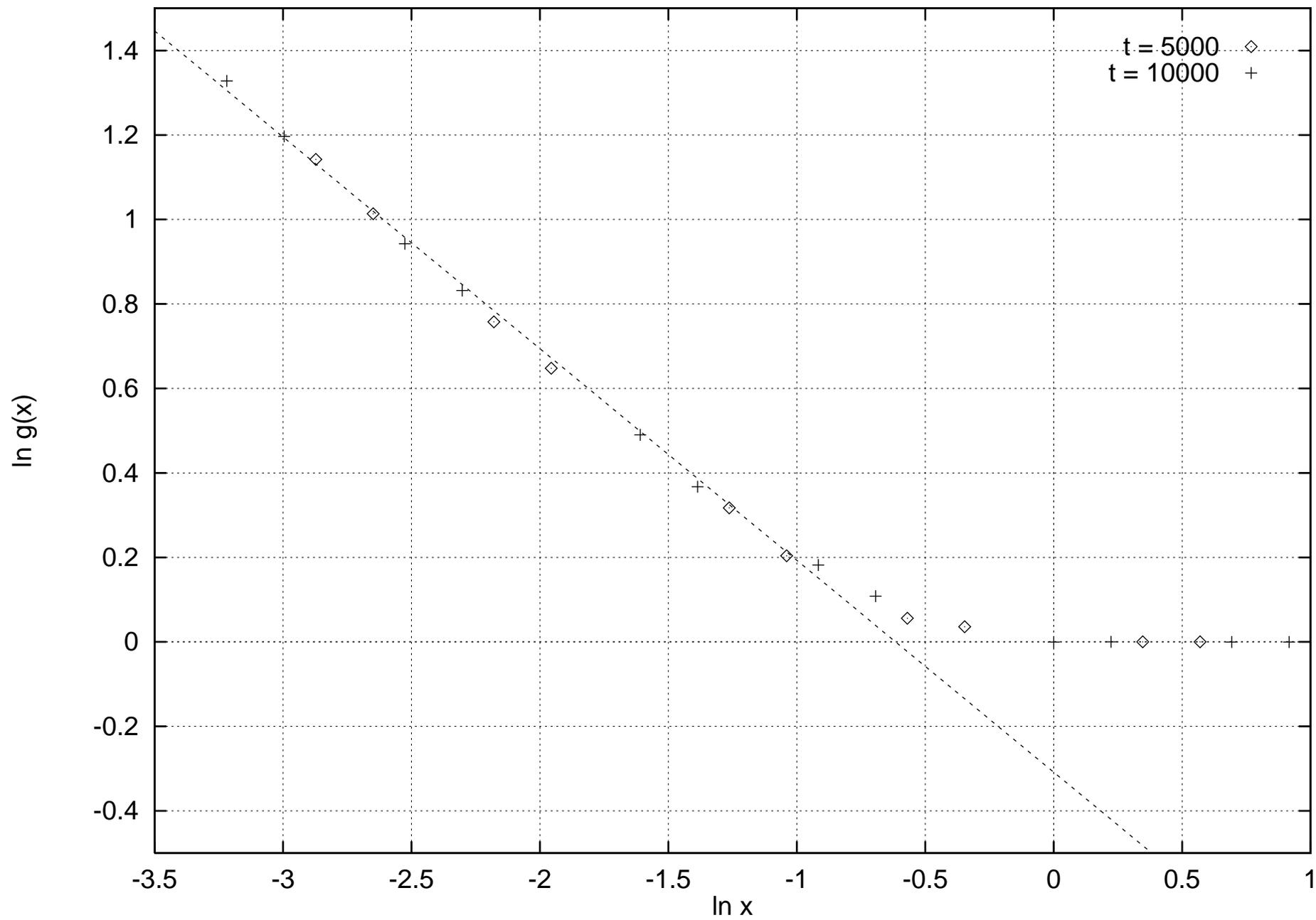
\end